\documentclass[12pt,article]{article}
\usepackage[german, english]{babel}
\usepackage[a4paper, left=2.5cm, right=2.5cm, top=3.5cm, bottom=3cm]{geometry}
\usepackage{amssymb}
\usepackage{amsmath}
\usepackage{amscd}
\usepackage{latexsym}
\usepackage{graphicx}
\usepackage{csquotes}
\usepackage{ifthen}
\usepackage{epsfig}
\usepackage{setspace}
\usepackage{hyperref}
\usepackage{dsfont}
\usepackage{mathrsfs}
\usepackage{tensor}
\usepackage{color}
\usepackage{cancel}
\usepackage{mathtools}
\usepackage{wrapfig}
\usepackage{float}
\usepackage{subfig}
\hypersetup{
 colorlinks=true,
 linkcolor=blue,
 citecolor=magenta,
}

\catcode`@=11
\renewcommand{\section}{\@startsection{section}{1}{0pt}{\medskipamount}
{\medskipamount}{\large\bf}}
\numberwithin{equation}{section}
\catcode`@=12

\newcommand{\N}{\mathds N}

\newcommand{\R}{\mathds R}
\newcommand{\C}{\mathds C}

\newcommand{\Acal}{{\mathcal A}}

\newcommand{\Hcal}{{\mathcal H}}
\newcommand{\Mcal}{{\mathcal M}}
\newcommand{\Ccal}{{\mathcal C}}
\newcommand{\Kcal}{{\mathcal K}}
\newcommand{\Ocal}{{\mathcal O}}
\newcommand{\Rcal}{{\mathcal R}}
\newcommand{\Lcal}{{\mathcal L}}
\newcommand{\Ical}{{\mathcal I}}

\newcommand{\Adot}{{\dot{\alpha}}}
\newcommand{\Bdot}{{\dot{\beta}}}
\newcommand{\adot}{{\dot{a}}}
\newcommand{\bdot}{{\dot{b}}}
\newcommand{\cDot}{{\dot{c}}}
\newcommand{\dDot}{{\dot{d}}}
\newcommand{\End}{\textrm{End}}
\newcommand{\CP}{\mathds{C}\mathrm{P}}
\newcommand{\ol}{1\textrm{loop}}
\newcommand{\eff}{\textrm{eff}}

\def\im{\mathrm{i}}
\def\ep{\mathrm{e}}
\def\pa{\partial}
\def\diff{\mathrm{d}}

\def\sfrac#1#2{{\textstyle\frac{#1}{#2}}}
\def\]{\right]}
\def\[{\left[}
\def\){\right)}
\def\({\left(}
\def\>{\rangle}
\def\<{\langle}
\def\+{\dagger}
\def\we{{\wedge}}
\def\={\ =\ }
\def\und{\quad\textrm{and}\quad}
\def\with{\quad\textrm{with}\quad}
\def\for{\quad\textrm{for}\quad}

\def\hs{\mathfrak{hs}}
\newcommand{\mso}{\mathfrak{so}}

\begin{document}




\begin{flushright}
 UWThPh-2023-28 \\
\end{flushright}
\hfill
\vskip 0.01\textheight
\begin{center}
{\Large\bfseries Modified Einstein equations from the $1$-loop \\[1ex] effective action of the IKKT model
}\\

\vskip 0.04\textheight

Kaushlendra Kumar\footnote{Address from July 1, 2023: Institute of Physics, Humboldt University, Zum Großen Windkanal 2, 12489 Berlin. Email: kumarkau@physik.hu-berlin.de}$^{a,b}$,
Harold C. Steinacker\,\footnote{Email: harold.steinacker@univie.ac.at}${}^{c}$
\vskip 0.03\textheight
$^{a}$ Institute of Theoretical Physics, Leibniz University Hannover \\
Appelstra{\ss}e 2, A-30167 Hannover, Germany \\[5pt]
$^{b}$ Erwin Schrödinger Institute for Mathematics and Physics \\ University of Vienna
\vskip 0.02\textheight
${}^{c}$ Department of Physics, University of Vienna, \\
Boltzmanngasse 5, A-1090 Vienna, Austria

\end{center}

\begin{abstract}
\noindent
We derive the equations of motion that arise from the one-loop effective action for the geometry of  3+1 dimensional quantum branes in the IKKT matrix model. These equations are cast into the form of generalized Einstein equations, with extra contributions from dilaton and axionic fields, as well as a novel anharmonicity tensor $C_{\mu\nu}$ capturing the classical Yang-Mills-type action. 
The resulting gravity theory approximately reduces to general relativity in some regime, but differs significantly at cosmic scales, leading to an asymptotically flat FLWR cosmological evolution governed by the classical action.

\end{abstract}

\thispagestyle{empty}
 \newpage

\section{Introduction}
\setcounter{page}{1} 
\noindent
The problem of embedding gravity in a consistent quantum theory is one of the big questions in theoretical physics. The accepted phenomenological description of gravity is given by general relativity (GR), where the geometry of space-time is related to matter via the Einstein equations. 
However, GR does not admit straightforward quantization, and it is plausible that the underlying quantum theory is based on very different degrees of freedom, such as strings or matrices. 

One specific proposal for such an underlying quantum theory is the IKKT matrix model, which is related to IIB string theory \cite{Ishibashi:1996xs,Aoki:1998vn}. As such, it can be viewed as a ``holographic" model for $9{+}1$ dimensional gravity on target space, leading to the standard issues of compactification and its inherent ambiguities, cf. \cite{Connes:1997cr,Klebanov:1997kv}. However, this model also admits a different, weakly coupled approach to $3{+}1$ dimensional gravity on suitable space-time branes via quantum effects. This approach is in the spirit of noncommutative gauge theory (cf. \cite{Aoki:1999vr,Szabo:2001kg,Douglas:2001ba,Rivelles:2002ez,Yang:2006hj,Szabo:2006wx,Steinacker:2019fcb,Kawai:2016wfh} and references therein), where target space has no direct physical meaning, and is not inhabited by propagating fields; it therefore does not need to be compactified. 

Within that framework, it was shown recently \cite{S22,S23} that the one-loop effective action indeed contains the Einstein-Hilbert action, supplemented by additional terms and degrees of freedom. This requires an extra compact ``fuzzy" factor $\Kcal_N$ in the underlying background brane configuration, which arises via the Higgs effect in the nonabelian sector of the theory.
However, the fundamental degrees of freedom are different from GR: 
The effective frame and metric are derived objects, which arise from {\em derivatives} of some underlying {\em gauge fields} given by the matrix background. 
Accordingly, the Yang-Mills-type matrix model action has two derivatives {\em less}  than the induced Einstein-Hilbert action. This is a very non-standard situation, and it is not evident how to obtain an intuitive set of equations that capture the combined one-loop effective action.
In particular, it does not suffice to consider the variation of the action w.r.t.~the frame or the metric, since those are neither fundamental nor independent.

The present paper is a first step to address this problem. We derive geometric equations of motion from the combined one-loop effective action computed in \cite{S23}, and cast them into a form as close to the Einstein equations as possible.
This is achieved by introducing a novel ``anharmonicity" tensor $C_{\mu\nu}$, which measures the deviation of the geometry from the Yang-Mills equations.
This can then be viewed as a novel term on the rhs of the effective Einstein-like equations.
The resulting gravity theory thus behaves similar to general relativity (apart from extra degrees of freedom) as long as 
$C_{\mu\nu}$ is negligible, but 
displays a rather different behavior at cosmic scales: In particular, it yields a non-standard evolution of FLRW space-times, which is rather insensitive to the detailed matter content. However, the full theory resulting from the present framework is very rich, and a satisfactory understanding of this theory will require much more work. 

This paper is organized as follows: Section 2 provides a short summary of the basic structure of the present approach, followed by the one-loop effective action in Section 3. The new results of this paper are given in Section 4, where we derive the equations of motion governing the geometry, and discuss some of its aspects. Some remaining open issues are discussed in Section 5, while several of the technical steps and supplementary comments are delegated to the appendix.

\section{Semi-classical geometry of the IKKT matrix model}
\noindent
The IKKT matrix model, defined in terms of bosonic matrices $T^\Adot$ with $\Adot = 0,\ldots,9$ and fermionic matrix-valued spinors $\Psi$, is given by the following action \cite{Ishibashi:1996xs}
\begin{equation}\label{action}
    S[T,\Psi] \= \frac{1}{g^2}\mathrm{Tr}\left( [T^\Adot,T^\Bdot][T_\Adot,T_\Bdot] 
    + \bar{\Psi}\Gamma_\Adot[T^\Adot,\Psi]\right)\ 
\end{equation}
where $g^2$ is a coupling constant. 
This model can be viewed as a constructive definition of critical IIB string theory with target space $\R^{1,9}$; here we simply consider it as the fundamental starting point.
The action admits several symmetries:
\begin{enumerate}
    \item[(a)] gauge invariance, i.e.~the model is invariant under the transformation $T^\Adot \rightarrow U^{-1}T^\Adot U$,
    \item[(b)] a global $SO(1,9)$ symmetry  acting on dotted indices with invariant tensor $\eta^{\Adot\Bdot}$, and
    \item[(c)] maximal supersymmetry.
\end{enumerate}
We will ignore the fermionic part in the following (apart from its contribution to the one-loop effective action), and focus on backgrounds defined by non-trivial matrix configurations $T^\Adot\in \End{(\Hcal)}$ acting on some Hilbert space $\Hcal$.
The classical equation of motion takes the form
\begin{equation}\label{eom}
    \square\, T_\Bdot \= [T^\Adot,[T_\Adot,T_\Bdot]] \= 0\ ,
\end{equation}
where $\square := [T^\Adot,[T_\Adot,\cdot]]$ is the matrix d'Alembertian. We will absorb the coupling constant $g^2$ in the matrices, since the self-interaction strength of the physical fluctuations becomes meaningful only on some given specific background; such interactions will not be important in the following. 

Many non-trivial matrix configurations $\bar{T}^\Adot$ -- and in particular 
solutions for the above eom \eqref{eom} -- can be interpreted 
 as (quantized) embedding maps of some brane configurations in the target space $\R^{1,9}$:
\begin{align}
   \bar T^\Adot\ : \ \ \Mcal \hookrightarrow \R^{1,9} \ .
   \label{background}
\end{align}
We restrict ourselves to the semi-classical regime in this paper, where matrices or operators can be interpreted as quantized functions on some underlying symplectic space.
This identification will be indicated by $\sim$ in the following. Then $\Box$ reduces to a second-order differential operator
\begin{align}
    \Box \sim - \{T^\Adot,\{T_\Adot,\cdot\}\} \ 
\end{align}
which will play the role of the d'Alembertian on suitable backgrounds.
We will study the geometrical equations of motion that determine such backgrounds at one loop, 
focusing on $3{+}1$ dimensional backgrounds 
defined by $\bar{T}^\adot$ with $\adot = 0,\ldots,3$, corresponding to 
   branes $\Mcal^{1,3}\hookrightarrow\R^{1,9}$. These branes are interpreted as a (quantized) space-time.  Fluctuations on such backgrounds are then parametrized as
\begin{equation}\label{fluc}
    T^\adot \= \bar{T}^\adot + \Acal^\adot\ ,
\end{equation}
which can be interpreted in terms of tangential gauge fields propagating along $\Mcal^{1,3}$, as well as transversal fluctuations interpreted in terms of scalar fields on the brane. 
We will consider not only such ``basic" $3{+}1$ dimensional branes with vanishing transversal fields, but also branes with non-vanishing VEV of the transversal fields, interpreted as compact fuzzy extra dimensions $\Kcal$. 
The latter is essential to obtain 
an induced Einstein-Hilbert action, and will be denoted as type II branes in section \ref{sec:background}.

\subsection{Effective metric, frame and torsion}
\noindent
The fluctuations of some background solutions $T^{\adot}$ \eqref{fluc} is governed by a universal effective metric $G^{\mu\nu}$, which arises from the kinetic term of the action\footnote{From now on $g^2$ is absorbed in the background and hence in $G_{\mu\nu}$.} \eqref{action} in the semi-classical limit $T^i\sim\phi$ \eqref{duality}:
\begin{equation}
    \textrm{Tr}\left([T^\adot,T^i][T_\adot,T_i]\right)\ \sim\ \int\diff^4x\sqrt{|G_{\mu\nu}|}\,G^{\mu\nu}\pa_\mu\phi\,\pa_\nu\phi\ ,
\end{equation}
where $|G_{\mu\nu}|$ denotes the metric determinant. This  effective metric $G_{\mu\nu}$ is suitably expressed in terms of a conformally related auxiliary metric $\gamma^{\mu\nu}$ 
as
\begin{equation}\label{confMetric}
    G_{\mu\nu} \= \rho^2\,\gamma_{\mu\nu}\ .
\end{equation}
Here $\gamma^{\mu\nu}$ is
given by 
\begin{equation}\label{auxMetric}
    \gamma^{\mu\nu}\ := \eta^{\adot\bdot}\,E^\mu_{\adot}\,E^\nu_{\bdot}\quad \with (\eta^{\adot\bdot}) \= \textrm{diag}(-1,1,1,1)\ ,   
\end{equation}
built from the frame $E^\mu_\adot$; the latter is associated with the background configuration $\bar{T}^\adot\sim t^\adot$ \eqref{duality}:
\begin{equation}\label{backFrame}
    E^\mu_{\adot} \= \{t_\adot,x^\mu\} 
\end{equation} 
where $x^\mu$ are local coordinates on $\Mcal^{1,3}$.
The corresponding coframe $E^\adot_\mu$ obeys following conditions\footnote{Notice here that the dotted/undotted labels are raised resp.~lowered with $\eta^{\adot\bdot}/\gamma^{\mu\nu}$ resp.~$\eta_{\adot\bdot}/\gamma_{\mu\nu}$.}
\begin{equation}
    E^\mu_\adot\,E^\bdot_\mu \= \delta^\bdot_\adot\quad \und\quad E^\mu_\adot\,E^\adot_\nu \= \delta^\mu_\nu\ .
\end{equation}
The conformal factor above is given by a dilaton field $\rho$,
which is determined in terms of the metric as follows (see Appendix \ref{VarTransf} for details),
\begin{equation} \label{rhoDef}
    \rho^2 \= \frac{\sqrt{|G_{\mu\nu}|}}{\rho_{_\Mcal}}\ 
    = \frac{\rho_{_\Mcal}}{|E_\adot^{\mu}|}\ .
\end{equation}
The density $\rho_{_\Mcal}$ arises from the volume form $\Omega = d^4x\, \rho_\Mcal$ of the symplectic (semi-classical) manifold, which relates the trace of operators $\End(\Hcal)\ni  \Phi \sim \phi$ \eqref{duality} to the integral via
\begin{equation}
    \textrm{Tr}(\Phi) \sim \int\limits_\Mcal\Omega\,\phi \= \int\limits_\Mcal\diff^4x\rho_{_\Mcal}\,\phi\ .
\end{equation}
The frame $E^\mu_{\Adot}$ turns out to obey a divergence constraint
\begin{equation}\label{constraint}
    \pa_\nu \(\rho_{_\Mcal}\,E^\nu_\adot\) \= 0\ ,
\end{equation}
which arises from the Jacobi identity. The existence of this constraint means that the frame $E^\mu_\adot$ does not admit local Lorentz (gauge) invariance. As a consequence, this frame has more physical content than in usual GR, such as an associated tensor two-form\footnote{In fact, the Jacobi identity for the torsion is a simple consequence of $T^\adot$ being exact, i.e. $\diff{T^\adot} = 0$.} $T^{\adot} := \diff{E^{\adot}}$ with $E^\adot {=} E^\adot_\mu\,\diff{x}^\mu$:
\begin{equation}
\label{T-dE}
    T^{\adot} \= \sfrac12 T_{\mu\nu}^{\ \ \adot}\,\diff{x}^\mu\we\diff{x}^\nu\ .
\end{equation}
The matrix model origin of this tensor can be attributed to the symplectic flow generated by $\theta^{\adot\bdot}$ \eqref{SympStruc}: 
\begin{equation}\label{torsion}
\begin{aligned}
    -\{\theta^{\adot\bdot},x^\mu\}\ =:\ T^{\adot\bdot\mu} &\= - E^{\adot\nu}\pa_\nu E^{\bdot\mu} + E^{\bdot\nu}\pa_\nu E^{\adot\mu}\ , \\
    T_{\mu\nu}^{\ \ \adot} &\= \pa_\mu E_\nu^{\adot} - \pa_\nu E_\mu^{\adot}\ 
    \end{aligned}
\end{equation}
using the Jacobi identity in the first line and raising/lowering indices with the frame to get the expression in the second line, which is compatible with \eqref{T-dE}. This tensor can also be understood geometrically as the torsion of the Weitzenböck connection $\nabla$ associated with the frame $E^\adot$, with connection coefficients $\Gamma_{\mu\nu}^{\ \ \sigma}$ given by 
\cite{S20,FS21} 
\begin{equation}
    \nabla_\mu E_{\adot}^\nu \= 0 \quad\implies\quad \Gamma^\sigma_{\ \mu\nu} \= -E_\nu^\adot\,\pa_\mu E_\adot^\sigma\ .
\end{equation}
The totally antisymmetric part of the torsion 
\begin{equation}\label{torsionAS}
    T^{\textrm{(AS)}\nu}_{\qquad\, \rho\mu} \= T^\nu_{\ \rho\mu} + T_{\mu\ \rho}^{\ \nu} + T_{\rho\mu}^{\ \ \nu} \quad\with\quad T_{\mu\nu}^{\ \ \rho} \= T_{\mu\nu}^{\ \ \adot}\,E^\rho_\adot\ 
\end{equation}
defines an ``axionic vector field" $T_\mu$  
via  Hodge-dualilty w.r.t the effective metric $G_{\mu\nu}$:
\begin{equation}
    T_\sigma\diff{x}^\sigma\ =:\ T \= -\star T^{(\textrm{AS})} \quad\for\quad T^{(\textrm{AS})} \= \frac{1}{3!}G_{\nu\nu'}T^{(\textrm{AS})\nu}_{\qquad \rho\mu}\,\diff{x}^{\nu'}\we\diff{x}^\rho\we\diff{x}^\mu\ .
\end{equation}
As a result, we have the following expression for $T_\sigma$,
\begin{equation}\label{defTa}
    T_{\sigma}\ :=\ -\frac{1}{3!}\sqrt{|G_{\mu\nu}|}\,\varepsilon_{\nu\mu\rho\sigma}G^{\mu\mu'}G^{\rho\rho'}\,T^{\textrm{(AS)}\nu}_{\qquad\mu'\rho'}\ .
\end{equation}
Then the equations of motion of the semi-classical matrix model imply that $T_\mu$ can be reduced to an
 {\it axion} field $\tilde{\rho}$ via \cite{FS21}
\begin{equation}\label{axion}
    \rho^2\,T_\mu \= \pa_\mu\tilde{\rho}\ .
\end{equation}

\subsection{Background geometry and Yang-Mills action}
\label{sec:background}

We will consider two basic types of backgrounds denoted by type I and type II branes.

\noindent
{\bf Brane I: $\Mcal^{1,3}\bar{\times}S^2_n$ as equivariant bundle.} An interesting class of solutions\footnote{Strictly speaking this is a solution of the classical (!) eom in the presence of a mass term, cf. \cite{Sperling:2019xar}. That issue will be addressed below at the one-loop level.} 
to eom \eqref{eom} is given by the following matrix configuration with {\it four} non-vanishing components:
\begin{equation}\label{I-T}
    \bar{T}^\adot \= \frac 1R\Mcal^{\adot 4}\ ,
    \qquad \adot = 0,...,3
\end{equation}
with the remaining $\bar {T}^i = 0$.
Here the matrices $\Mcal^{\adot 4}$ are part of the $\mathrm{SO}(2,4)$ generators, $\Mcal^{ab}\in\End(\Hcal_n)$ with $a,b=0,\ldots,5$, that belong to unitary irreducible representations $\Hcal_n$ 
 known as the {\em doubleton series} and are labeled by $n\in\N$. One can then
 identify the quantum space of operators $\End(\Hcal_n)$ on the associated Hilbert space
 with 
 the classical space of functions $\Ccal(\CP^{1,2})$  \cite{Sperling:2019xar}:
\begin{equation}\label{duality}
    \Ccal(\CP^{1,2})\ \cong\ \End(\Hcal_n)\ \ .
\end{equation}
The underlying $6$-dimensional symplectic manifold $\CP^{1,2}$ turns out to be an $\mathrm{SO}(3,1)$-equivariant\footnote{The local stabilizer $SO(3)$ at each point of $\Mcal^{1,3}$ acts non-trivially on the $S^2$-fiber.} bundle $\CP^{1,2}$ over the spacetime $\Mcal^{1,3}$ with fuzzy $2$-spheres $S^2_n$ acting as fibres, known as twistor space. 
 This leads to a structure denoted as (semi-classical) {\em covariant quantum space},
\begin{equation}
    \CP^{1,2}\ \overset{loc}{\cong}\ \Mcal^{1,3}\times S^2_n\ ,
\end{equation}
where $\Mcal^{1,3}$ is generated by Cartesian coordinates $x^\mu$, while $S^2_n$ is generated by generators $t^\mu$, which also serve as momentum generators and define the undeformed semi-classical background matrices $T^\adot \sim t^\adot$ \cite{Sperling:2019xar}.
The semi-classical regime arises in the limit $n\rightarrow \infty$, where the matrix commutators reduce to Poisson brackets $[\cdot,\cdot] \rightarrow \im\{\cdot,\cdot \}$ as follows:
\begin{equation}\label{SympStruc}
     \Theta^{\adot\bdot}\ := -i\ [T^\adot,T^\bdot]\ \sim\ \{t^\adot,t^\bdot\} \ =:\ \theta^{\adot\bdot}\ .
\end{equation}
$\Mcal^{1,3}$ also acquires an effective metric describing a cosmological FLRW background with negative spatial curvature and a global length scale set by $R$. The latter should be determined dynamically using the 1-loop effective action, or simply by introducing a mass parameter in the model.

\noindent
Explicitly, the effective metric and the dilaton on $\Mcal^{1,3}$ are found to be \cite{Sperling:2019xar}
\begin{align}
    G_{\mu\nu} &= \sinh(\eta) \eta_{\mu\nu}, \qquad 
    \rho^2 = \sinh^3(\eta)
    \label{cosm-metric-rho} 
\end{align}
in Cartesian coordinates, which corresponds to an FLRW space-time with scale parameter 
\begin{align}
\label{a-rho-BG}
    a(t) \sim \rho \  \sim \frac 32 t
\end{align}
at late times. More details about this cosmological space-time can be found in \cite{Battista:2022hqn}.


\paragraph{Fluctuations.}

Now consider the fluctuations on $\Mcal^{1,3}$.
A generic fluctuation $\Acal^\mu$ \eqref{fluc} can be expressed in terms of functions $\Acal\sim \phi$ on $\C P^{1,2}$ via the duality \eqref{duality}, expanded into  $S^2$ harmonics $Y_{sm}$ as
\begin{equation}
    \phi \= \sum\limits_{s=0}^{n}\sum\limits_{m=-s}^{s} \phi_{sm}(x)\,Y_{sm}\ .
\end{equation}
The spin $s$ here are eigenmodes of the $S^2_n$-Laplacian arising from the decomposition of $\Box$ \eqref{eom}:
\begin{equation}
    \Box \= \Box_{\Mcal^{1,3}} + m_s^2\ , \with m_s^2 = \frac{3s}{R^2}\ .
\end{equation}
This analysis prompts one to study the higher spin gauge theory through the splitting of the algebra, 
\begin{equation}
    \Ccal \= \bigoplus\limits_{s\geq 0} \Ccal^s\ ,
\end{equation}
into higher-spin modules $\Ccal^s$. We will largely ignore this higher spin part in the following, except for their contribution to the one-loop computation.

\noindent
{\bf Brane II: $\Mcal^{1,3}\bar{\times}S^2_n\times\Kcal$ with compact $\Kcal$.}

To obtain induced gravity, we also consider 
another matrix configuration $T^{\Adot}$ with an extra compact brane $\Kcal$ embedded in the transversal dimensions of target space $\R^{1,9}$.
This arises from the background given by 
\begin{equation}\label{II-Ta}
    T^\adot \sim x^\adot: \quad \Mcal^{1,3} \hookrightarrow \R^{1,3}\ , \qquad \adot = 0,...,3
\end{equation}
describing the quantum spacetime $\Mcal^{1,3}$
embedded along the first $3{+}1$ directions, and some compact quantum space $\Kcal$ (such as a fuzzy sphere) embedded along the remaining $6$ directions as
\begin{equation}\label{II-Ti}
    T^i \sim z^i: \quad \Kcal \hookrightarrow \R^6\ ,\qquad i=4,\ldots,9\ .
\end{equation}
The exact structure of $\Kcal$ is not relevant for our discussion except for the finite, discrete (positive) spectrum, labeled by $\Lambda$, of its Laplacian $\Box_{\Kcal}$ arising from the splitting $\Box = \Box_{\Mcal^{1,3}}+\Box_{\Kcal}$ \eqref{eom}:
\begin{equation}\label{KKmass}
    \Box_{\Kcal}\,\lambda_{\Lambda} \= m_{\Lambda}^2\,\lambda_{\Lambda}\ ,\quad m_{\Lambda}^2 \= m_{\Kcal}^2\mu_{\Lambda}^2
\end{equation}
associated with eigenmodes $\lambda_{\Lambda}\in \End(\Hcal_{\Kcal})$\footnote{Here $\mu_\Lambda^2$ refers to the tower of discrete KK modes, while $m_\Kcal^2$ is some overall scale associated with the compact $\Kcal$.}. The matrices then act on the product   Hilbert space  $\Hcal = \Hcal_{\Mcal^{1,3}}\otimes \Hcal_{\Kcal}$,
where $\Hcal_{\Kcal}$ is finite-dimensional.
We can  expand generic fluctuations $\Acal\sim \phi \in \End(\Hcal)$ in term of the Kaluza--Klein (KK) modes as
\begin{equation}
    \phi \= \sum\limits_{\Lambda}\phi_{\Lambda}\,\lambda_{\Lambda} \ 
\end{equation}
where $\phi_{\Lambda}\in \End(\Hcal_{\Mcal^{1,3}})$.
This product background induces a mass for the KK modes $\phi_\Lambda$ on space-time:
\begin{equation}\label{KKsplit}
    \Box\phi_\Lambda \= (\Box_{\Mcal} + m_{\Lambda}^2)\phi_\Lambda\ .
\end{equation}

\paragraph{The semi-classical Yang-Mills action.}
The bosonic part of the matrix model action \eqref{action} reduces in the semi-classical limit for the above $3{+}1$ dimensional backgrounds $T^\adot$ to the 
following  Yang-Mills-type action
\begin{equation}\label{}
    S_{\rm YM} =  \mathrm{Tr}[T^\Adot,T^\Bdot][T_\Adot,T_\Bdot] 
    \sim  -\int\limits_\Mcal \Omega\,
 \{T^\adot,T^\bdot\}\{T_\adot,T_\bdot\} \ .
\label{YMterm}
\end{equation}
For type II background branes, the transversal matrices contribute an extra term to this action given by
\begin{align}
\label{YMterm-extra}
 S_{\rm YM}^{(\Kcal)} 
 \sim -\int\limits_\Mcal \Omega\,
 \{T^i,T^j\}\{T_i,T_j\} \= -\int\limits_\Mcal \Omega\, m_\Kcal^4 F_\Kcal^2
\end{align}
where $F_\Kcal^2$ is some discrete number depending on the structure of $\Kcal$.
This term will play an important role in stabilizing the vacuum.
There is also a mixed term 
$\int \Omega\{T^\adot,T^j\}\{T_\adot,T_j\}$, which 
amounts to a kinetic term for $\Kcal$. That term is
expected to suppress possible variations of $m_\Kcal$, but will be omitted in the following.

\section{The one-loop action}
\noindent
The $1$-loop effective action $\Gamma_{\ol}$ is defined via the Gaussian integral around a given matrix configuration $T^\adot$:
\begin{equation}\label{partFn}
    Z_{\ol}[T] \= \int\limits_{\ol}\diff{T}\diff{\Psi}\diff{c}\diff{\bar{c}}\,\ep^{\im S[T,\Psi,c]} 
     = \ep^{S[T]+\Gamma_{\ol}[T]}
\end{equation}
with fermion $\Psi$ and ghost $c$ (for gauge-fixing) contributions integrated out and the bare action \eqref{action} evaluated on the background under consideration.
This path integral can be regularised by a Feynman $\im\varepsilon$ term through the following addition,
\begin{equation}
    S \rightarrow S+ \im\varepsilon\,\textrm{Tr}\sum_\Adot(T^\Adot)^2\ .
\end{equation}
Due to the maximal supersymmetry of our matrix model the first three terms of the effective action cancel, resulting in the following non-trivial contribution (dropping higher-order contributions $\Ocal(\Box^{-5})$)
\begin{align}
    \Gamma_{\ol;4} &= \frac{\im}{8}\mathrm{Tr}\bigg(\Big(\Box^{-1}M_{\Adot\Bdot}^{(V)}[\Theta^{\Adot\Bdot},\cdot]\Big)^4 - \frac{1}{2}\Big(\Box^{-1}M_{\Adot\Bdot}^{(\psi)}[\Theta^{\Adot\Bdot},\cdot]\Big)^4 \bigg)
  = \frac{3\im}{4}\textrm{Tr}\(\frac{V_4}{(\Box-\im\varepsilon)^4}\) 
\end{align}
where $M_{\Adot\Bdot}$ are $\textrm{SO}(1,9)$ generators acting on vector $(V)$ and spinor $(\psi)$ representations, 
the trace is over $\End(\Hcal)$,
and 
\begin{equation}
    V_4 \= -4\textrm{tr}(\delta\Theta^4)+(\textrm{tr}(\delta\Theta^2))^2\ 
\end{equation}
with $\delta \Theta = [\Theta,.]$.
Recently, this $4$th order $1$-loop term has been computed \cite{S22,S23} for the brane-II configuration (\ref{II-Ta}-\ref{II-Ti}), observing that the traces are UV finite. On such product space $\Mcal\times\Kcal$, 
the various contributions from $V_4$ split into three parts $V_4=V_4^\Mcal+V_4^\Kcal+V_4^{\Mcal-\Kcal}$ with contributions coming from $\Mcal$, or $\Kcal$, or both:
\begin{equation}
\begin{aligned}
    V_4^\Mcal &\= (\delta\Theta^{\adot\bdot}\delta\Theta_{\adot\bdot})(\delta\Theta^{\cDot\dDot}\delta\Theta_{\cDot\dDot}) - 4(\delta\Theta^{\adot\bdot}\delta\Theta_{\bdot\cDot}\delta\Theta^{\cDot\dDot}\delta\Theta_{\dDot\adot})\ ,\\
    V_4^\Kcal &\= (\delta\Theta^{ij}\delta\Theta_{ij})(\delta\Theta^{kl}\delta\Theta_{kl}) - 4(\delta\Theta^{ij}\delta\Theta_{jk}\delta\Theta^{kl}\delta\Theta_{li})\ ,\\
    V_4^{\Mcal-\Kcal} &\= 2(\delta\Theta^{\adot\bdot}\delta\Theta_{\adot\bdot})(\delta\Theta^{ij}\delta\Theta_{ij})\ .
\end{aligned}
\end{equation}
The mixed term $V_4^{\Mcal-\Kcal}$ is most important\footnote{Note that in all these computations, the frame and all derived quantities are strictly speaking higher-spin valued. We will neglect such components in the present paper, lacking an appropriate formalism to treat these components.} for us, as it yields an Einstein-Hilbert-like contribution to the action in terms of the effective metric $G_{\mu\nu}$. The trace can be evaluated in the semi-classical regime using the above geometric quantities, notably the torsion \eqref{torsion} which arises from 
$\delta \Theta \sim i\{\Theta,.\}$.
This leads to
\begin{equation}\label{effAction}
    \Gamma^{\Mcal-\Kcal}_{\ol;4} \= \int\limits_\Mcal\diff^4x\,\frac{\sqrt{|G|}}{16\pi G_N}\(\Rcal + \frac{1}{2}T\cdot T - 2\rho^{-2}\pa\rho\cdot\pa\rho + 2\rho^{-1}G_N^{-1}\pa\rho\cdot\pa G_N\)\ 
\end{equation}
 (here the dot-product is w.r.t.~the effective metric, e.g.~$T\cdot T = T_\mu T_\nu G^{\mu\nu}$), using
 the following identity \cite[Appendix E]{FS21}
\begin{equation}\label{offShell}
    \Rcal \= -\frac{1}{2}T^\mu_{\ \sigma\rho}T_{\mu\sigma'}^{\quad\rho}G^{\sigma\sigma'} - \frac{1}{2}T_{\mu}T_{\nu}G^{\mu\nu} + 2\rho^{-2}G^{\mu\nu}\pa_\mu\rho\,\pa_\nu\rho - 2\nabla^\mu_{(G)}\(\rho^{-1}\pa_\mu\rho\)\ .
\end{equation}
Here $G_N$ plays the role of Newton's constant and is given  by
\begin{equation}\label{GNdef}
    G_N \= \frac{\pi^3\rho^2}{2c_\Kcal^2m_\Kcal^2}\ 
\end{equation}
in terms of the scale $m_\Kcal$
and a constant $c_\Kcal\gg 1$.
In addition, the one-loop effective action also contains contributions (from $V_4^\Kcal$) that are naturally interpreted as vacuum energy. They
have the form \cite{S23} 
\begin{align}
\label{S-vac}
    S_{\rm vac} = \int\limits_\Mcal \Omega \rho^{-2}
    \Big(C_1 m_\Kcal^4 + C_2 \frac{1}{R^4} 
    + C_3 \frac{1}{R^8 m_\Kcal^4} \Big) \ 
\end{align}
with large but finite constants $C_i$.
In the traditional approach to gravity, the vacuum energy contributions would have the form $\int\sqrt{G}\, \Lambda^4$, which amounts to a cosmological constant. The presence of the rigid symplectic volume form $\Omega$ in the present framework suggests that the usual cosmological constant problem should not arise here. 
Moreover, the induced vacuum energy density decreases with the cosmic expansion due to
$\Omega \rho^{-2} \sim \sqrt{|G|} \rho^{-4}$ and $\rho \sim a(t)$.
However, this vacuum energy will play a role in determining $m_\Kcal$.

\section{Emergent gravity from the one-loop effective action}
\noindent
Now we consider the full one-loop effective action, comprising the bare Yang-Mills matrix model \eqref{YMterm} and the 1-loop contribution $\Gamma_{\ol}$ \eqref{effAction}:
\begin{align}
    S_\eff  &=   S_{\rm YM} + S_{\mathrm{matter}} + \Gamma_{\ol}   \ .
\end{align}
We have also included explicitly 
a (classical) matter action 
\begin{align}
    S_{\mathrm{matter}} = \int\diff^4x\,\sqrt{|G_{\mu\nu}|}\,\Lcal_{\mathrm{matter}} \ ,
\end{align}
 which arises from the fermionic terms of the matrix model, 
as well as contributions of the
non-abelian sector arising from the fuzzy extra dimensions $\Kcal$, depending on the specific background.
To understand the equations of motion (eom) for this coupled system, we consider a generic background fluctuation $\delta T^\adot = \Acal^\adot$ on $\Mcal^{1,3}$ \eqref{fluc}, which yields a frame fluctuation around $E_\adot^\mu$ \eqref{backFrame}:
\begin{equation}
\label{frame-perturbations}
    \delta E_\adot^\mu \equiv \delta_\Acal E_\adot^\mu \= \{\Acal_\adot,x^\mu\}\ .
\end{equation}
For computational simplicity we first consider the variation in the metric $\delta G_{\mu\nu}$, in the dilaton $\delta\rho$ and in the KK mass $\delta m_\Kcal$ \eqref{KKmass} as independent\footnote{The variation in the first two can be written in terms of $\delta E^\mu_\adot$ as we will see below, while $m_\Kcal$ is a priori an independent degree of freedom.}, so that we obtain
\begin{equation}\label{varScoupled}
    \delta S_\eff \= 4\int\limits_\Mcal\diff^4x\rho_{_\Mcal}\,\Box T_\adot\,\delta T^\adot - \frac{1}{2}\int\limits_\Mcal\diff^4x\sqrt{|G_{\mu\nu}|}\,T^{(m)}_{\mu\nu}\delta G^{\mu\nu} - 4\int\limits_\Mcal \Omega\,F_\Kcal^2 m_\Kcal^3\delta m_\Kcal + \delta \Gamma_{\ol}
\end{equation}
where $T^{(m)}_{\mu\nu}$ is the standard stress-energy tensor for matter which is given by
\begin{equation}\label{Tmunu}
    T^{(m)}_{\mu\nu} \= -\frac{2}{\sqrt{|G_{\mu\nu}|}}\frac{\delta S_{\mathrm{matter}}}{\delta G^{\mu\nu}} \= -2\frac{\delta \Lcal_{\textrm{matter}}}{\delta G^{\mu\nu}} + G_{\mu\nu}\Lcal_{\textrm{matter}}\ .
\end{equation}
The variation of the $1$-loop term, i.e.~$\delta \Gamma_{\ol}$, is obtained after a cumbersome exercise involving variations of its individual components (see Appendix \ref{details}); the result \eqref{delGamma} can be neatly expressed using
\begin{equation}
    \rho \= \ep^\sigma \quad\und\quad m_\Kcal \= \ep^\lambda\ ,
\end{equation}
to rewrite the logarithmic derivations. We obtain (including the vacuum energy contributions)
\begin{equation}{\small
\label{delGamma2}
\begin{aligned}
    \delta \Gamma_{\ol} &= \int\limits_\Mcal\diff^4x\frac{\sqrt{|G|}}{16\pi G_N}\Big\{\Rcal_{\mu\nu}-\sfrac{1}{2}G_{\mu\nu}\Rcal - \sfrac12\big(T_\mu T_\nu-\sfrac12G_{\mu\nu}T\cdot T\big) - \big(2\pa_\mu\sigma\pa_\nu\sigma-3G_{\mu\nu}\pa\sigma\cdot\pa\sigma\big) \\
    &\hspace{0.25\linewidth}+ 2\big(2\pa_\mu\sigma\pa_\nu\lambda-3G_{\mu\nu}\pa\sigma\cdot\pa\lambda\big) - 4\big(\pa_\mu\lambda\pa_\nu\lambda-G_{\mu\nu}\pa\lambda\cdot\pa\lambda\big)\\
    &\hspace{0.35\linewidth}+ 2\big(\pa_\mu\pa_\nu\sigma-G_{\mu\nu}\Box_G\sigma\big) - 2\big(\pa_\mu\pa_\nu\lambda-G_{\mu\nu}\Box_G\lambda\big)\Big\} \delta G^{\mu\nu}\\
    &+\int\limits_\Mcal\diff^4x\frac{\sqrt{|G|}}{16\pi G_N}\Big\{{-2}\Rcal - T\cdot T + 4\pa\sigma\cdot\pa\sigma - 8\pa\sigma\cdot\pa\lambda + 8\pa\lambda\cdot\pa\lambda - 4\Box_G\sigma + 4\Box_G\lambda \Big\}\delta\sigma \\
    &+\int\limits_\Mcal\diff^4x\frac{\sqrt{|G|}}{16\pi G_N}\Big(2\Rcal + T\cdot T - 4\pa\sigma\cdot\pa\sigma + 4\Box_G\sigma \Big)\delta\lambda -2 \int\limits_\Mcal \Omega \rho^{-2}
    \Big(C_1 m_\Kcal^4 
    + \frac{C_2}{R^4}
    + \frac{C_3}{R^8 m_\Kcal^4} \Big) \delta\sigma \\
    &- \int\limits_\Mcal\diff^4x\frac{\rho^2}{16\pi G_N}\varepsilon^{\nu\sigma\mu\kappa}\Big\{T_\kappa T_{\nu\mu}^{\ \ \adot} + m_\Kcal^{-2}E^\adot_{\mu}\pa_\nu(m_\Kcal^2T_\kappa)\Big\}\delta E_{\adot\sigma} + 4 \int\limits_\Mcal \Omega \rho^{-2}
    \Big(C_1 m_\Kcal^4 
    - \frac{C_3}{R^8 m_\Kcal^4} \Big)\delta\lambda\ .
\end{aligned}
}
\end{equation}
We can further simplify this expression by diagonalizing (most of) the kinetic terms via
\begin{equation}
    \tilde{\sigma} \= \sigma - \lambda\ ,
\end{equation}
to obtain the following result for $\delta\Gamma_{\ol}$ \eqref{delGamma2}:
\begin{equation}
{\small
\label{delGamma2b}
\begin{aligned}
    \delta \Gamma_{\ol} \= &\int\limits_\Mcal\diff^4x\frac{\sqrt{|G|}}{16\pi G_N}\bigg[\Big\{\Rcal_{\mu\nu}-\sfrac{1}{2}G_{\mu\nu}\Rcal - \sfrac12\big(T_\mu T_\nu-\sfrac12G_{\mu\nu}T\cdot T\big) - \big(2\pa_\mu\tilde{\sigma}\pa_\nu\tilde{\sigma}-3G_{\mu\nu}\pa\tilde{\sigma}\cdot\pa\tilde{\sigma}\big)\\
    &\qquad\qquad\qquad- \big(2\pa_\mu\lambda\pa_\nu\lambda-G_{\mu\nu}\pa\lambda\cdot\pa\lambda\big)+ 2\big(\pa_\mu\pa_\nu\tilde{\sigma}-G_{\mu\nu}\Box_G\tilde{\sigma}\big) \Big\} \delta G^{\mu\nu}\\
    &\quad\quad+ \Big({-2}\Rcal - T\cdot T + 4\pa\tilde{\sigma}\cdot\pa\tilde{\sigma} + 4\pa\lambda\cdot\pa\lambda - 4\Box_G\tilde{\sigma}\Big)\delta\tilde{\sigma} + \Big({-} 8\pa\tilde{\sigma}\cdot\pa\lambda + 4\Box_G\lambda \Big)\delta\lambda\bigg]\\
    &- \int\limits_\Mcal\diff^4x\frac{\rho^2}{16\pi G_N}\varepsilon^{\nu\sigma\mu\kappa}\Big\{T_\kappa T_{\nu\mu}^{\ \ \adot} + m_\Kcal^{-2}E^\adot_{\mu}\pa_\nu(m_\Kcal^2T_\kappa)\Big\}\delta E_{\adot\sigma}\\
    &- 2\int\limits_\Mcal \Omega \rho^{-2}
    \Big(C_1 m_\Kcal^4 
    + \frac{C_2}{R^4}
    + \frac{C_3}{R^8 m_\Kcal^4} \Big) \delta\tilde{\sigma} + 2\int\limits_\Mcal \Omega \rho^{-2}
    \Big(C_1 m_\Kcal^4 
    - \frac{C_2}{R^4}
    - 3\frac{C_3}{R^8 m_\Kcal^4} \Big) \delta\lambda \ .
\end{aligned}
}
\end{equation}
We obtain the eom for $m_\Kcal$ corresponding to $\delta\lambda$ in \eqref{delGamma2} using \eqref{rhoDef} and also including the bare matrix model contribution $S_{\rm YM}$ \eqref{varScoupled} as follows:
\begin{equation}\label{eomMass}
    \Rcal + \frac{1}{2}T\cdot T - 2\pa\sigma\cdot\pa\sigma + 2\Box_G\sigma 
    - 32\pi G_N \rho^{-2}F_\Kcal^2 m_\Kcal^4 + 32\pi G_N \rho^{-4}
    \big( C_1 m_\Kcal^4 
    - C_3 \frac{1}{R^8 m_\Kcal^4}\big)
    \ = 0\ .
\end{equation}
This equation determines $m_\Kcal$. To be specific, we assume that the vacuum energy contributions given by the last terms 
 with $C_1, C_3>0$
dominate. Then $m^2_\Kcal$ is given as the minimum of the effective potential,
\begin{align}
\label{m-K-min-oneloop}
 m_\Kcal^8 
    \approx \frac{C_3}{C_1} \frac{1}{R^8} \ .
\end{align}
This provides a dynamical stabilization of $\Kcal$ through quantum effects; accordingly, we will assume that $\lambda = const$ from now on, and drop the tilde on $\sigma$. Together with
\eqref{rhoDef}, the above expression for $\delta\Gamma_{\ol}$ \eqref{delGamma2} takes a simpler form: 
\begin{equation}\label{varSeff1}
{\small
\begin{aligned}
    \delta \Gamma_{\ol} \= &\int\limits_\Mcal\Omega\frac{\rho^2}{16\pi G_N}\bigg[\Big\{\Rcal_{\mu\nu}-\sfrac{1}{2}G_{\mu\nu}\Rcal - \sfrac12\big(T_\mu T_\nu - \sfrac12G_{\mu\nu}T\cdot T\big) - \big(2\pa_\mu\sigma\pa_\nu\sigma-3G_{\mu\nu}\pa\sigma\cdot\pa\sigma\big)\\
    &\hspace{0.03\linewidth}+ 2\big(\pa_\mu\pa_\nu\sigma - G_{\mu\nu}\Box_G\sigma\big)\Big\} \delta G^{\mu\nu} - \rho_{_\Mcal}^{-1}\varepsilon^{\nu\sigma\mu\kappa}\(T_\kappa T_{\nu\mu}^{\ \ \adot}+E^\adot_{\mu}\pa_\nu T_\kappa\)\delta E_{\adot\sigma}\\
    &\hspace{0.03\linewidth}- 2\Big(\Rcal + \sfrac12T\cdot T - 2\pa\sigma\cdot\pa\sigma + 2\Box_G\sigma + 16\pi G_N \rho^{-4}
    \big(C_1 m_\Kcal^4 + \frac{C_2}{R^4} 
    + \frac{C_3}{R^8 m_\Kcal^4}\big)  \Big)\delta\sigma\bigg]\ .
\end{aligned}
}
\end{equation}
The variations $\delta G^{\mu\nu}$ and $\delta\sigma$ can be expressed as follows (Appendix \ref{VarTransf}):
\begin{align}
    \delta\sigma &\= \frac{1}{2} E_{\adot\sigma}\delta E^{\adot\sigma}\ ,\label{varSigma}\\
    \delta G^{\mu\nu} &\= 2\rho^{-2}E^\mu_\adot\delta E^{\adot\nu} -G^{\mu\nu}E_{\adot\sigma}\delta E^{\adot\sigma}\ .\label{varGmunu}
\end{align}
Plugging these relations 
into \eqref{varSeff1} while also using the constraint arising from the $m_\Kcal$ eom \eqref{eomMass} to eliminate $\Rcal$ gives us the following variation of $\Gamma_{\ol}$ in terms of the frame\footnote{We have adjusted some indices with effective/auxiliary metric while using \eqref{confMetric}.}:
\begin{equation}\label{varSeff2}
\begin{aligned}
    \delta \Gamma_{\ol} \= &\int\limits_\Mcal\frac{\Omega}{8\pi G_N}\bigg[\Big\{\Rcal_{\mu\lambda} - \sfrac12T_\mu T_\lambda - 2\big(\pa_\mu\sigma\pa_\lambda\sigma+G_{\mu\lambda}\pa\sigma\cdot\pa\sigma\big)+ 2\big(\pa_\mu\pa_\lambda\sigma+\sfrac12G_{\mu\lambda}\Box_G\sigma\big)\\
    &\hspace{0.2\linewidth}- 8\pi G_N\rho^{-4}G_{\mu\lambda}\big(2\rho^2F_{\Kcal}^2m_\Kcal^4-C_1m_\Kcal^4+\frac{C_2}{R^4}+3\frac{C_3}{R^8m_\Kcal^4}\big)\Big\}E^{\mu}_\adot\\
    &\hspace{0.36\linewidth} - \sfrac12\rho_{_\Mcal}^{-1}G_{\sigma\lambda}\varepsilon^{\nu\sigma\mu\kappa}\(T_\kappa T_{\nu\mu\adot} + E_{\adot\mu}\pa_\nu T_\kappa\)
    \bigg]\delta E^{\adot\lambda}\ .
\end{aligned}
\end{equation}
We can also express the variation of the matter action in terms of the frame as
\begin{align}
    \delta S_{\textrm{matter}} 
    &\= - \frac{1}{2}\int\limits_\Mcal\diff^4x\sqrt{|G_{\mu\nu}|}\,T^{(m)}_{\mu\nu}\delta G^{\mu\nu} \= - \frac{1}{2}\int\limits_\Mcal\Omega\,\rho^2\,T^{(m)}_{\mu\nu}\(2\rho^{-2}E^\mu_\adot\delta E^{\adot\nu} + G^{\mu\nu}E^{\adot\sigma}\delta E_{\adot\sigma}\) \nonumber\\
    &\= - \frac{1}{2}\int\limits_\Mcal\Omega\(2T^{(m)}_{\mu\lambda}E^\mu_\adot\delta E^{\adot\lambda}  - \rho^2\,T^{(m)}E^{\adot}_\lambda\delta E_{\adot}^\lambda\)\nonumber\\
    &\= - \frac{1}{2}\int\limits_\Mcal\Omega\(2T^{(m)}_{\mu\lambda}E^\mu_\adot -  G_{\sigma\lambda}T^{(m)}E^{\sigma}_\adot\)\delta E^{\adot\lambda}\ ,
    \label{varMatter}
\end{align}
where  $T^{(m)}=G^{\mu\nu}T^{(m)}_{\mu\nu}$ \eqref{Tmunu} is the trace of the matter stress-energy tensor.
This extra trace contribution arises due to the conformal factor $\rho^2$ in the effective metric, but will drop out in the equations of motion.

Using the above result \eqref{varSeff2} and \eqref{varMatter},
the variation of the full action for the coupled system \eqref{varScoupled} takes the following form  
{\small 
\begin{align}
    \delta S_\eff \= &\int\limits_\Mcal\frac{\Omega}{8\pi G_N}\bigg[\Big\{\Rcal_{\mu\lambda} - \sfrac12T_\mu T_\lambda - 2\big(\pa_\mu\sigma\pa_\lambda\sigma+G_{\mu\lambda}\pa\sigma\cdot\pa\sigma\big)+ 2\big(\pa_\mu\pa_\lambda\sigma+\sfrac12G_{\mu\lambda}\Box_G\sigma\big) \nonumber\\
    &-8\pi G_N\big(T^{(m)}_{\mu\lambda}-\sfrac12G_{\mu\lambda}T^{(m)}\big)- 8\pi G_N\rho^{-4}G_{\mu\lambda}\big(2\rho^2F_{\Kcal}^2m_\Kcal^4-C_1m_\Kcal^4+\frac{C_2}{R^4}+3\frac{C_3}{R^8m_\Kcal^4}\big)\Big\}E^{\mu}_\adot \nonumber\\
    & - \sfrac12\rho_{_\Mcal}^{-1}G_{\sigma\lambda}\varepsilon^{\nu\sigma\mu\kappa}\(T_\kappa T_{\nu\mu\adot} + E_{\adot\mu}\pa_\nu T_\kappa\)
    \bigg]\delta E^{\adot\lambda}\ + 4\int\limits_\Mcal\Omega\,\(\Box T^\adot\)\delta T_\adot\ .
    \label{Seff}
\end{align}
}
 Notice that $\rho_{_\Mcal}^{-1}\varepsilon^{\nu\sigma\mu\kappa}$  in the third line is indeed a tensor.


To make the resulting eom more transparent, we can find  
\begin{align}
    C_{\adot\mu} = C_{\nu\mu} E^\nu_\adot
\end{align}
such that 
\begin{align}
    \Box T_\adot \= \{C_{\adot\mu},x^\mu\} \ .
    \label{Camu-def-BG}
\end{align}
This can generically be solved by the ansatz $C_{\adot\mu} = \{B_\adot,x_\mu\}$, which leads to
\begin{align}
    \Box_x B_\adot = \Box T_\adot \sim \rho^2 \Box_G T_\adot \ , \quad \mbox{where} \quad
    \Box_x = -\{x_\mu,\{x^\mu,.\}\} \ .
    \label{B-C-equation}
\end{align}
Such a $C_{\adot\mu}$  is not unique, but admits the following ``gauge invariance"
\begin{align}
 C_{\adot\mu} \to C_{\adot\mu} + \partial_\mu C_\adot   \ .
 \label{C-ambiguity}
\end{align}
For example, for the unperturbed background\footnote{This follows from the relations $\{t^\adot,x^\mu\} =  \sinh(\eta)\eta^{\adot\mu}$ and $\Box t^\adot = \frac{3}{R^2} t^\adot$ and $\{\sinh(\eta),x^\mu\} = r^2 t^\mu$ on $\Mcal^{1,3}$. 
Here $r$ is a scale parameter fixed by the underlying representation of $\mso(4,2)$, see \cite{Sperling:2019xar}.} $\Mcal^{1,3}$ we have
\begin{align}
\label{Camu-BG}
   C_{\adot\mu} = 
\frac 3{r^2 R^2} \sinh(\eta) \eta_{\adot\mu} 
\end{align}
with $B_\adot = \frac{3}{r^2 R^2} t_\adot$. 
More generally, $C_{\mu\nu}$ measures the ``anharmonicity" of the background, i.e. the deviation from $\Box T^\adot = 0$.
We can then write the variation of the bare YM action in the form
\begin{equation}
\begin{aligned}
    \delta S_{\rm YM} &\= 4\int\limits_\Mcal\diff^4x\rho_{_\Mcal}\, \Box T_\adot\,\delta T^\adot 
     \sim 4\int\limits_\Mcal\diff^4x\rho_{_\Mcal}\,
     \{C_{\adot\mu},x^\mu\} \,\delta T^\adot\\
     &= - 4\int\limits_\Mcal\diff^4x\rho_{_\Mcal}\,
     C_{\adot\mu} \{\delta T^\adot,x^\mu\}\\
     &= - 4\int\limits_\Mcal\diff^4x\rho_{_\Mcal}\,
     C_{\adot\mu} \delta E^{\adot\mu} \ .
\end{aligned}
\end{equation}
This is consistent with the ambiguity \eqref{C-ambiguity} due to the divergence constraint \eqref{constraint}. That ambiguity will eventually be fixed in Appendix \ref{sec:div-einstein}.

Assuming for the moment that the fluctuations of the frame are unconstrained, this leads to the 
 eom 
{\small 
\begin{align}
 & \Big\{\Rcal_{\mu\lambda} - 2\big(\pa_\mu\sigma\pa_\lambda\sigma+G_{\mu\lambda}\pa\sigma\cdot\pa\sigma\big)+ 2\big(\pa_\mu\pa_\lambda\sigma+\sfrac12G_{\mu\lambda}\Box_G\sigma\big) - \sfrac12T_\mu T_\lambda \nonumber \\
  &\ - 8\pi G_N\rho^{-4}G_{\mu\lambda}\big(2\rho^2F_{\Kcal}^2m_\Kcal^4-C_1m_\Kcal^4+\frac{C_2}{R^4}+3\frac{C_3}{R^8m_\Kcal^4}\big)
  \Big\}E^{\mu}_\adot - \sfrac12\rho_{_\Mcal}^{-1}G_{\sigma\lambda}\varepsilon^{\nu\sigma\mu\kappa}\(T_\kappa T_{\nu\mu\adot} + E_{\adot\mu}\pa_\nu T_\kappa\) \nonumber\\ 
   &\quad = {8\pi G_N}\bigg[T^{(m)}_{\mu\lambda} - \frac 12 G_{\mu\lambda}T^{(m)}
    + 4 C_{\mu\lambda}\bigg] E^\mu_\adot\ .
    \label{Einstein-e-mod1}
\end{align}
}
This can be written without the frame as follows:
{\small 
\begin{align}
    \Rcal_{\mu\lambda}
    &= {8\pi G_N}\bigg[T^{(m)}_{\mu\lambda} - \frac 12 G_{\mu\lambda}T^{(m)}
    + \rho^{-4}G_{\mu\lambda}\big(2\rho^2F_{\Kcal}^2m_\Kcal^4-C_1m_\Kcal^4+\frac{C_2}{R^4}+3\frac{C_3}{R^8m_\Kcal^4}\big) + 4 C_{\mu\lambda} \bigg] \nonumber\\
    & \quad + 2\big(\pa_\mu\sigma\pa_\lambda\sigma{+}G_{\mu\lambda}\pa\sigma\cdot\pa\sigma\big)- 2\big(\pa_\mu\pa_\lambda\sigma{+}\sfrac12G_{\mu\lambda}\Box_G\sigma\big) \nonumber\\[1ex]
   & \quad  + \sfrac12T_\mu T_\lambda + \sfrac12\rho_{_\Mcal}^{-1}G_{\sigma\lambda}\varepsilon^{\nu\sigma\rho\kappa}\(T_\kappa \tensor{T}{_\nu_\rho_\mu} {+} \gamma_{\mu\rho} \pa_\nu T_\kappa\)
   \label{Einstein-e-mod2}
\end{align}
}
which in the absence of totally antisymmetric torsion $T^{\rm (AS)} = 0$ simplify as 
\begin{equation}
\label{Einstein-e-mod2a}
\begin{aligned}
    \Rcal_{\mu\lambda}
    &= {8\pi G_N}\bigg[T^{(m)}_{\mu\lambda} - \frac 12 G_{\mu\lambda}T^{(m)} +\rho^{-4}G_{\mu\lambda}\big(2\rho^2F_{\Kcal}^2m_\Kcal^4 -C_1m_\Kcal^4+\frac{C_2}{R^4}+3\frac{C_3}{R^8m_\Kcal^4}\big) + 4C_{\mu\lambda}\bigg]\\
    &\quad + 2\big(\pa_\mu\sigma\pa_\lambda\sigma+G_{\mu\lambda}\pa\sigma\cdot\pa\sigma\big)- 2\big(\pa_\mu\pa_\lambda\sigma+\sfrac12G_{\mu\lambda}\Box_G\sigma\big)\ .  
\end{aligned}
\end{equation}
In particular, this implies the trace relation
{\small
\begin{align}
    \Rcal
    &\= {8\pi G_N}\bigg[-  T^{(m)}
    + 4\rho^{-4}\big(2\rho^2F_{\Kcal}^2m_\Kcal^4
    -C_1m_\Kcal^4+\frac{C_2}{R^4}+3\frac{C_3}{R^8m_\Kcal^4}\big) 
    + 4 C\bigg] 
    + 10\pa\sigma\cdot\pa\sigma 
    - 6 \Box_G\sigma  
    \label{Einstein-e-mod2-trace}
\end{align}
}
where $C = G^{\mu\lambda}C_{\mu\lambda}$. Combining these, 
we obtain the {\em modified Einstein equations}
{\small
\begin{align}
\boxed{
\begin{aligned}
    \Rcal_{\mu\lambda} - \sfrac 12 G_{\mu\lambda}\Rcal
    &=  {8\pi G_N}\bigg[T^{(m)}_{\mu\lambda} 
    - \frac 1{\rho^4}G_{\mu\lambda}\big(2\rho^2F_{\Kcal}^2m_\Kcal^4
    -C_1m_\Kcal^4+\frac{C_2}{R^4}+\frac{3C_3}{R^8m_\Kcal^4}\big) 
    + 4 (C_{\mu\lambda} - \sfrac 12 G_{\mu\lambda} C)\bigg]  \\
    & \quad + 2\big(\pa_\mu\sigma\pa_\lambda\sigma
    -\pa_\mu\pa_\lambda\sigma
    + G_{\mu\lambda}(\Box_G\sigma
    -\sfrac 32\pa\sigma\cdot\pa\sigma)
    \big)\ .  
    \end{aligned}
    }
\label{Einstein-e-mod3}
\end{align}
}
This may need to be amended by the axionic field $T_\mu$ as above.
We note that the anharmonicity tensor $C_{\mu\lambda}$, which arises from the  Yang--Mills term via \eqref{Camu-def-BG},
plays the role of a source term in the Einstein equation. However, remember that  $C_{\mu\lambda}$ is only determined up to a total divergence $E^\adot_\lambda \partial_\mu C_\adot $  \eqref{C-ambiguity}, which reflects the divergence constraint of the frame.
Thus the variations $\delta E_{\adot\mu}$ are not independent, so that
 the above equations only hold up to precisely this ambiguity. 
This can be indicated by writing them as 
\begin{equation}
\label{Einstein-e-mod4}
\begin{aligned}
    \Rcal_{\mu\lambda} - \frac12G_{\mu\lambda} \Rcal
    &\= {8\pi G_N}\bigg[T^{(m)}_{\mu\lambda}
    -\rho^{-4}G_{\mu\lambda}\big(2\rho^2F_{\Kcal}^2m_\Kcal^4
    -C_1m_\Kcal^4+\frac{C_2}{R^4}+3\frac{C_3}{R^8m_\Kcal^4}\big) \\
    &\quad + 4 \Big(C_{\mu\lambda} + 
    E^\adot_\lambda\partial_\mu C_\adot
    -\frac 12 G_{\mu\lambda}(C + \rho^{-2} E^{\adot\mu}\partial_\mu C_\adot)\Big)\bigg] \\
    &\quad     + 2\big(\pa_\mu\sigma\pa_\lambda\sigma 
    -\pa_\mu\pa_\lambda\sigma 
    + G_{\mu\lambda}(\Box_G\sigma
    -\frac 32\pa\sigma\cdot\pa\sigma)\big)
\end{aligned}
\end{equation}
for some undetermined $C_\adot$, and similarly in the version including the axionic field $T_\mu$.
This  ambiguity will be fixed in Appendix \ref{sec:div-einstein} through the 
conservation of the Einstein tensor.

In the above computations,
it should be kept in mind that the
geometric quantities generically contain non-vanishing higher spin components, which may not be fully accounted for in the above form. Nevertheless, the above equations should be satisfied
 at least in the weak gravity regime
for the classical, dominant components of the geometry.

\paragraph{Cosmological FLRW background.}

Now consider the 
undeformed FLRW background $\Mcal^{1,3}$. Then the anharmonicity tensor $C_{\mu\lambda}$ arising from the classical YM action  is given explicitly by 
\begin{align}
    C_{\mu\lambda} &\= \frac 3{r^2 R^2} \sinh(\eta) \eta_{\adot\mu} E^\adot_\lambda
    =  \frac {3}{r^2 R^2} \eta_{\mu\lambda} 
    =  \frac {3}{r^2 R^2\sinh(\eta)}  G_{\mu\lambda} 
\end{align}
where $G_{\mu\nu}$
is the effective metric \eqref{cosm-metric-rho} on $\Mcal^{1,3}$.
This acts like an additional vacuum energy contribution to the gravitational equations.
At early times $\cosh(\eta) = O(1)$, the vacuum energy contributions determined by $C_i$ which scale like $\rho^{-4} G_{\mu\nu}$ dominate the 
classical YM contribution $C_{\mu\nu} \sim 1/r^2 R^2\, \eta_{\mu\nu}$
for sufficiently large\footnote{Note that $C_i$ are 
typically very large sums of structure constants associated to $\Kcal$, cf. \cite{S23}.}
$C_{i}$.  We  assume for simplicity that the
contribution from $F_\Kcal^2$ is sub-leading and can be dropped.
Then both sides of \eqref{Einstein-e-mod3}
behave like $\rho^{-2} G_{\mu\nu} + \rho^{-4} \tau_\mu \tau_\nu$ (noting \eqref{GNdef} and \eqref{a-rho-BG}, where $\tau$ is the time-like FLRW vector field \cite{Steinacker:2020xph});
moreover, the fuzzy extra dimensions $\Kcal$ are stabilized via \eqref{m-K-min-oneloop}.
All this 
supports the consistency of the present background $\Mcal^{1,3} \times \Kcal$ at early times.
 This is confirmed in a more careful study of the 1-loop equations of motion for this background in \cite{Battista:2023glw}.

\paragraph{Late-time regime.}

At late times $\eta \gg 1$,
 the vacuum energy described by $C_{i}$  decreases with the cosmic expansion as $\rho^{-4} \sim a(t)^{-4}$. Hence the vacuum energy contribution to the curvature becomes negligible, so that no cosmological constant problem arises. 
On the other hand, the $C_{\mu\lambda}$ tensor from the  
undeformed background $\Mcal^{1,3}$ clearly has a significant impact on the 1-loop equation of motion, so that the background is no longer consistent. But this is easily understood and fixed: 
the undeformed background $T^\adot$ is a solution of the classical IKKT model only in the presence of an extra mass term in the action. That mass term was put in by hand to stabilize   $\Mcal^{1,3}$  at the classical level; it should be removed in the quantum theory.
Upon dropping this mass term, that background is modified to the following classical solution 
\begin{align}
    \tilde T^\adot \sim \alpha(\eta) t^\adot
\end{align}
with $\alpha(\eta) \sim e^\eta$, as shown in   \cite{Battista:2023glw}.
This leads to the
effective metric $\tilde G_{\mu\nu} = \cosh^3(\eta)\eta_{\mu\nu}$, 
which describes a somewhat modified  $k=-1$ FLRW geometry rather similar to $\Mcal^{1,3}$, with an accelerated expansion and cosmic scale parameter $a(t) \sim \frac 52 t$. Then the above background tensor vanishes 
\begin{align}
 \tilde C_{\mu\nu} = 0\ 
\end{align}
since $\tilde{\Box} \tilde T^\adot = 0$.
In the presence of local perturbations, some anharmonicity $C_{\mu\nu} \neq 0$ will arise, which is related to the geometry in a non-local way that remains to be understood in detail.
However, note that $C_{\mu\nu}$ is determined via \eqref{Camu-def-BG} as a 
``first derivative" (in a non-local sense) of the background $T^\adot$, just like the frame $E^{\adot\mu}$, and is therefore soft compared
with the curvature. This reflects the fact that
the Einstein-Hilbert action has two more derivatives than the bare matrix-model action which determines $C_{\mu\nu}$.
Moreover, $C_{\mu\nu}$ is expected to vanish in vacuum in the linearized regime, where
perturbations are harmonic (hence $C_{\mu\nu}=0$) and Ricci-flat \cite{Steinacker:2016vgf}.
Therefore $C_{\mu\nu}$ is expected to be subleading compared with the standard GR contributions, at least at shorter scales. 
On the other hand, the scaling argument does not apply for long scales, where any non-vanishing $C_{\mu\nu}$ may have significant impact.
A more careful examination of the cosmological evolution at one loop \cite{Battista:2023glw} suggests that the classical equations of motion indeed
dominate over the induced one-loop effects, leading to a picture that is rather different from GR and less sensitive to the detailed matter content\footnote{At the classical level, matter does not act as a source for the curvature in the matrix model.}.

All this suggests that the present framework leads to a physically reasonable emergent gravity theory, which is close to general relativity (extended by dilatonic and axionic contributions\footnote{Note that the dilaton and axion are subject to constraints in the present framework \cite{FS21}, and will behave differently from more standard extensions of GR. }) at shorter scales, but deviates significantly from GR on cosmic scales.

\section{Discussion and conclusion}
\noindent
The main result of this paper is an equation for the geometry of brane solutions of the IKKT matrix model, based on the one-loop effective action obtained in \cite{S22,S23}. This is non-trivial because the fundamental degrees of freedom of the matrix model are given by (fluctuations of) the matrix background, while the frame and metric are derived objects, given by certain {\em derivatives} of the matrix fluctuations. That is not surprising in a Yang--Mills theory, but it makes it hard to cast the equations into a recognizable form that can be compared with general relativity. We managed to find a transparent form for such a {\em modified Einstein equation}, in terms of a novel ``anharmonicity" tensor $C_{\mu\nu}$ which measures the deviation from a harmonic background.

This result is clearly a useful step towards understanding the physics of the gravity theory which emerges from the matrix model. In particular, it provides strong evidence that the resulting gravity theory behaves similarly to GR in some regime, which is essential for its physical viability.
However, we have not yet found a useful way to describe the effect of the new tensor $C_{\mu\nu}$ explicitly, which is related to the metric geometry in a non-local way. This is perhaps the most pressing task to be addressed in follow-up work.

There are many other open questions that need to be addressed in future work. 
In particular, it turns out that 
perturbations \eqref{frame-perturbations} of the present type of covariant quantum spaces 
$\Mcal^{1,3}$
generically involve higher-spin ($\hs$)-valued components,
 which mirror the geometric ones. This implies that local perturbations of the torsion are mirrored by $\hs$-valued components, which have been dropped in this paper. 
These carry essentially the same information and are not expected to significantly change the above results, except possibly in the extreme IR regime\footnote{See \cite{Battista:2023glw} for an analysis of cosmological solutions including these $\hs$ components.}. That issue needs to be addressed in detail elsewhere.

Finally, we recall that the fuzzy extra dimensions $\Kcal$ lead to an interesting nonabelian gauge theory in the present framework, through spontaneous breaking of some internal gauge group $U(n)$ \cite{CSZ11}. This can lead to an interesting low-energy gauge theory with (approximate) fermionic zero modes governed by the effective metric discussed in this paper.

\section*{Acknowledgements}
\noindent
KK is grateful to the Erwin Schr{\" o}dinger Institute at Vienna for supporting this work through a Junior Research Fellowship in 2023.
The work of HS is supported by the 
Austrian Science Fund (FWF) grants P32086 and P36479.

\appendix

\section{Variation of $\Gamma_{\ol}$} \label{details}
\noindent
To compute the variation of the $1$-loop term $\Gamma_{\ol}$ \eqref{effAction} we first decompose it into following four parts:
\begin{equation}\label{1loopSplit}
    \Gamma_{\ol} \= S_\Rcal + S_T + S_\rho + S_{G_N}\ ,
\end{equation}
where the different components of the action are as follows:
\begin{equation}
\begin{aligned}
    S_\Rcal &\= \int\limits_\Mcal\diff^4x\,\frac{\sqrt{|G|}}{16\pi G_N}\Rcal\ ,\\
    S_\rho &\= -\int\limits_\Mcal\diff^4x\,\frac{\sqrt{|G|}}{16\pi G_N}\frac{2}{\rho^2}\pa\rho\cdot\pa\rho\ ,
\end{aligned}
\qquad
\begin{aligned}
    S_T &\= \int\limits_\Mcal\diff^4x\,\frac{\sqrt{|G|}}{16\pi G_N} \frac{1}{2}T\cdot T\ ,\\
    S_{G_N} &\=  \int\limits_\Mcal\diff^4x\,\frac{\sqrt{|G|}}{16\pi G_N}\frac{2}{\rho\,G_N}\pa\rho\cdot\pa G_N\ .
\end{aligned}
\end{equation}
In the following calculations, we will be working in Riemann normal coordinates where the first derivatives of the metric vanishes. This means that we will routinely interchange partial derivatives with covariant derivatives, since they differ by Christoffel symbols that are also vanishing.

\noindent
{\bf Variation of $S_\Rcal$.} Using $\Rcal = G^{\mu\nu}\Rcal_{\mu\nu}$ and the well-known variation of the determinant:
\begin{equation}\label{varDet}
    \delta |G_{\mu\nu}| \= |G_{\mu\nu}|G^{\mu\nu}\delta G_{\mu\nu}  \= -|G_{\mu\nu}|G_{\mu\nu}\delta G^{\mu\nu}\ ,
\end{equation}
we can express the variation of $S_\Rcal$ in \eqref{1loopSplit} as follows:
\begin{equation}\label{delSr1}
    \delta S_\Rcal \= \int\limits_\Mcal\diff^4x\frac{\sqrt{|G|}}{16\pi G_N}\Big\{\big(\Rcal_{\mu\nu}-\sfrac{1}{2}G_{\mu\nu}\Rcal\big)\delta G^{\mu\nu} - \Rcal\,G_N^{-1}\delta G_N + G^{\mu\nu}\delta\Rcal_{\mu\nu}\Big\}\ .
\end{equation}
Here we have made use of the following result:
\begin{align}
    G_N^{-1}\delta G_N \= 2(\rho^{-1}\delta \rho - m_\Kcal^{-1} \delta m_\Kcal)\ .
\end{align}
The non-trivial term here is the last one involving $\delta\Rcal_{\mu\nu}$, i.e.
\begin{equation}\label{intI1}
    \Ical_1 \= \int\limits_\Mcal\diff^4x\frac{\sqrt{|G|}}{16\pi G_N}G^{\mu\nu}\delta\Rcal_{\mu\nu}\ .
\end{equation}
A well-known geometrical result gives us $\delta\Rcal_{\mu\nu}$ in terms of the variation of Christoffel symbol:
\begin{equation}\label{deltaR}
    \delta\Rcal_{\mu\nu} \= \nabla_\rho\delta\Gamma^\rho_{\ \mu\nu} - \nabla_\nu\delta\Gamma^\rho_{\ \rho\mu}\ .
\end{equation}
We also have the following result in the normal coordinates:
\begin{equation}
    \delta\Gamma^\rho_{\ \mu\nu} \= \frac{1}{2}G^{\rho\kappa}\(\nabla_\mu\delta G_{\nu\kappa} + \nabla_\nu\delta G_{\mu\kappa} - \nabla_\kappa\delta G_{\mu\nu}\)\ .
\end{equation}
Equipped with this result, we can now compute the first term of \eqref{deltaR} to get
\begin{equation}
\begin{aligned}
    G^{\mu\nu}\nabla_\rho\delta\Gamma^\rho_{\ \mu\nu} &\= \frac{1}{2}\nabla_\rho\Big[G^{\mu\nu}\big\{-\nabla_\mu\(G_{\nu\kappa}\delta G^{\rho\kappa}\)-\nabla_\nu\(G_{\mu\kappa}\delta G^{\rho\kappa}\)\big\} + G^{\rho\kappa}\nabla_\kappa\(G_{\mu\nu}\delta G^{\mu\nu}\)\Big] \\
    &\= -\nabla_\rho\nabla_\kappa\delta G^{\rho\kappa} + \frac{1}{2}G_{\mu\nu}\Box_G\delta G^{\mu\nu}\ ,
\end{aligned}
\end{equation}
while the other term of \eqref{deltaR} can be computed by noting the following contraction,
\begin{equation}
    \delta\Gamma^\rho_{\ \rho\mu} \= \frac{1}{2}G^{\rho\kappa}\nabla_\mu\delta G_{\rho\kappa}\ ,
\end{equation}
to obtain
\begin{equation}\label{deltaRfinal}
    G^{\mu\nu}\nabla_\nu\delta\Gamma^\rho_{\ \rho\mu} \= -\frac{1}{2}G_{\rho\kappa}\square_G\delta G^{\rho\kappa}\ .
\end{equation}
Taken together, these relations (\ref{deltaR}-\ref{deltaRfinal}) gives us the required integral,
which we partially integrate to obtain (up to some boundary term)
\begin{equation}\label{intI1final}
    \Ical_1 \= \frac{c_\Kcal^2}{8\pi^4}\int\limits_\Mcal\diff^4x\sqrt{|G_{\mu\nu}|}K_{\mu\nu}\delta G^{\mu\nu}  \with K_{\mu\nu} \= G_{\mu\nu}\square_G\(\frac{m_\Kcal^2}{\rho^2}\) -\pa_\mu\pa_\nu\(\frac{m_\Kcal^2}{\rho^2}\)\ .
\end{equation}
A straightforward computation then gives us the following bulky expression for this integrand:
\begin{equation}\label{intIK}
\begin{aligned}
    K_{\mu\nu}\delta G^{\mu\nu} &\= \Big\{{-}\frac{2}{\rho^2}\(\pa_\mu m_\Kcal\,\pa_\nu m_\Kcal - G_{\mu\nu}\pa m_\Kcal\cdot\pa m_\Kcal\) + \frac{8m_\Kcal}{\rho^3}\(\pa_\mu m_\Kcal\pa_\nu\rho - G_{\mu\nu}\pa m_\Kcal\cdot\pa\rho\) \\
    &\qquad\qquad- \frac{6m_\Kcal^2}{\rho^4}\(\pa_\mu\rho\,\pa_\nu\rho-G_{\mu\nu}\pa\rho\cdot\pa\rho\) - \frac{2m_\Kcal}{\rho^2}\(\pa_\mu\pa_\nu m_\Kcal-G_{\mu\nu}\square_G m_\Kcal\) \\
    &\qquad\qquad\qquad\qquad\qquad\qquad\qquad\qquad\qquad+ \frac{2m_\Kcal^2}{\rho^3}\(\pa_\mu\pa_\nu\rho-G_{\mu\nu}\square_G\rho\)\!\Big\}\delta G^{\mu\nu}\ .   
\end{aligned}
\end{equation}
Plugging these results (\ref{intI1final}-\ref{intIK}) back into \eqref{delSr1} we obtain
{\small
\begin{equation}\label{delSr2}
\begin{aligned}
    \delta S_\Rcal \= \int\limits_\Mcal\diff^4x&\frac{\sqrt{|G|}}{16\pi G_N}\bigg[\Big\{(\Rcal_{\mu\nu}-\sfrac{1}{2}G_{\mu\nu}\Rcal) - \frac{2}{m_\Kcal^2}\(\pa_\mu m_\Kcal\,\pa_\nu m_\Kcal - G_{\mu\nu}\pa m_\Kcal\cdot\pa m_\Kcal\) \\
    & + \frac{8}{m_\Kcal\rho}\(\pa_\mu m_\Kcal\pa_\nu\rho - G_{\mu\nu}\pa m_\Kcal\cdot\pa\rho\) - \frac{6}{\rho^2}\(\pa_\mu\rho\,\pa_\nu\rho-G_{\mu\nu}\pa\rho\cdot\pa\rho\) \\
    & - \frac{2}{m_\Kcal}\(\pa_\mu\pa_\nu m_\Kcal-G_{\mu\nu}\square_G m_\Kcal\) + \frac{2}{\rho}\(\pa_\mu\pa_\nu\rho-G_{\mu\nu}\square_G\rho\)\Big\}\delta G^{\mu\nu}-\Rcal G_N^{-1}\delta G_N\bigg]\ .
\end{aligned}
\end{equation}
}

\noindent
{\bf Variation of $S_T$.} First of all, a straightforward computation involving \eqref{varDet} yields
\begin{equation}\label{delSt1}
    \delta S_T \= \int\limits_\Mcal\diff^4x\frac{\sqrt{|G|}}{16\pi G_N}\Big\{\sfrac12\big(T_\mu T_\nu - \sfrac12G_{\mu\nu}T\cdot T\big)\delta G^{\mu\nu} - \sfrac12 T\cdot T\,G_N^{-1}\delta G_N + G^{\mu\nu}T_\mu\delta T_\nu \Big\}\ .
\end{equation}
Furthermore, the term involving $\delta T_\nu$ simplifies as follows (see \cite[Appendix E]{FS21}):
\begin{equation}\label{intI2}
\begin{aligned}
    \Ical_2 &\= \int\limits_\Mcal\diff^4x\frac{\sqrt{|G|}}{16\pi G_N}G^{\mu\nu}T_\mu\delta T_\nu\ ,\\
    &\= \frac{1}{2}\int\limits_\Mcal\diff^4x\frac{\sqrt{|G|}}{16\pi G_N}G^{\lambda\rho}T_\lambda\Big\{\!\sqrt{|G_{\mu\nu}|}^{-1}\rho^2\big(G_{\rho\kappa}\varepsilon^{\nu\sigma\mu\kappa}\delta T_{\nu\sigma\mu}+\varepsilon^{\nu\sigma\mu\kappa}T_{\nu\sigma\mu}\delta G_{\rho\kappa}\big) \\
    &\hspace{0.55\linewidth}+ \big(\sfrac{2}{\rho}\delta\rho+\sfrac{1}{2}G_{\mu\nu}\delta G^{\mu\nu}\big)2T_\rho\!\Big\}\\
    &\= \frac{1}{2}\int\limits_\Mcal\diff^4x\frac{\sqrt{|G|}}{16\pi G_N}\Big\{\!-2\big(T_\mu T_\nu-\sfrac{1}{2}G_{\mu\nu}T\cdot T\big)\delta G^{\mu\nu} + \sfrac{4}{\rho}T\cdot T\delta\rho\!\Big\}\\
    &\hspace{0.5\linewidth}+\frac{c_\Kcal^2}{16\pi^4}\int\limits_\Mcal\diff^4x\,m_\Kcal^2\varepsilon^{\nu\sigma\mu\kappa}T_\kappa\delta T_{\nu\sigma\mu}\ ,
\end{aligned}
\end{equation}
where in the last line we have employed the following definition of $T_\rho$ which is equivalent to \eqref{defTa}
\begin{equation}
    T_\rho \= \frac{1}{2}\sqrt{|G_{\mu\nu}|}^{-1}\rho^2G_{\rho\kappa}\varepsilon^{\nu\sigma\mu\kappa}T_{\nu\sigma\mu}\ .
\end{equation}
The last part of this integral can be computed as follows:
\begin{align}
    \Ical_3 &= \frac{c_\Kcal^2}{16\pi^4}\int\limits_\Mcal\diff^4x\,m_\Kcal^2\varepsilon^{\nu\sigma\mu\kappa}T_\kappa\delta T_{\nu\sigma\mu} \nonumber\\
    &= \frac{c_\Kcal^2}{16\pi^4}\int\limits_\Mcal\diff^4x\,m_\Kcal^2T_\kappa\varepsilon^{\nu\sigma\mu\kappa}\Big\{\!T_{\nu\sigma}^{\ \ \adot}\delta E_{\adot\mu} + 2E_{\adot\mu}\pa_\nu\delta E^\adot_{\sigma} \!\Big\}\nonumber\\
    &= \frac{c_\Kcal^2}{16\pi^4}\int\limits_\Mcal\diff^4x\,m_\Kcal^2T_\kappa\varepsilon^{\nu\sigma\mu\kappa}T_{\nu\sigma}^{\ \ \adot}\delta E_{\adot\mu} -\frac{c_\Kcal^2}{16\pi^4}\int\limits_\Mcal\diff^4x\,\varepsilon^{\nu\sigma\mu\kappa}\Big\{\!2E^\adot_{\mu}\pa_\nu(m_\Kcal^2T_\kappa) + m_\Kcal^2T_\kappa T_{\nu\mu}^{\ \ \adot}\!\Big\}\delta E_{\adot\sigma}\nonumber\\
    &= -\frac{c_\Kcal^2}{8\pi^4}\int\limits_\Mcal\diff^4x\,\varepsilon^{\nu\sigma\mu\kappa}\Big\{\!m_\Kcal^2T_\kappa T_{\nu\mu}^{\ \ \adot} + E^\adot_{\mu}\pa_\nu(m_\Kcal^2T_\kappa)\!\Big\}\delta E_{\adot\sigma}
    \label{intI3}
\end{align}
where we have used the definition of the torsion \eqref{torsion} and \eqref{torsionAS} in the second step and integration by parts in the third one. Putting together \eqref{delSt1}, \eqref{intI2} and \eqref{intI3} we obtain
\begin{align}
    \delta S_T &\= \int\limits_\Mcal\diff^4x\frac{\sqrt{|G|}}{16\pi G_N}\Big\{{-}\sfrac12\big(T_\mu T_\nu - \sfrac12G_{\mu\nu}T\cdot T\big)\delta G^{\mu\nu} - \sfrac12 T\cdot T\,G_N^{-1}\delta G_N\Big\} \nonumber\\
    &- \int\limits_\Mcal\diff^4x\frac{\rho^2}{16\pi G_N}\varepsilon^{\nu\sigma\mu\kappa}\Big\{T_\kappa T_{\nu\mu}^{\ \ \adot} + m_\Kcal^{-2}E^\adot_{\mu}\pa_\nu(m_\Kcal^2T_\kappa)\Big\}\delta E_{\adot\sigma}\ .
    \label{delSt2}
\end{align}

\noindent
{\bf Variation of $S_\rho$ and $S_{G_N}$.} We can compute the variation of $S_\rho$ as follows,
{\small
\begin{equation}\label{delSrho}
\begin{aligned}
    \delta S_\rho &\= -2\int\limits_\Mcal\diff^4x\frac{\sqrt{|G|}}{16\pi G_N}\Big\{\rho^{-2}\big(\pa_\mu\rho\pa_\nu\rho-\sfrac12 G_{\mu\nu}\pa\rho\cdot\pa\rho\big)\delta G^{\mu\nu} - \rho^{-2}\pa\rho\cdot\pa\rho \big(G_N^{-1}\delta G_N+2\rho^{-1}\delta\rho\big) \\
    &\hspace{0.7\linewidth}+ 2\rho^{-2}\pa\rho\cdot\pa\delta\rho \Big\}\\
    &\= -2\int\limits_\Mcal\diff^4x\frac{\sqrt{|G|}}{16\pi G_N}\Big\{\rho^{-2}\big(\pa_\mu\rho\pa_\nu\rho-\sfrac12 G_{\mu\nu}\pa\rho\cdot\pa\rho\big)\delta G^{\mu\nu} - \rho^{-2}\pa\rho\cdot\pa\rho \big(G_N^{-1}\delta G_N+2\rho^{-1}\delta\rho\big) \\
    &\hspace{0.4\linewidth} -2\big(\sfrac{2}{\rho^2m_\Kcal}\pa\rho\cdot\pa m_\Kcal-\sfrac{4}{\rho^3}\pa\rho\cdot\pa\rho + \rho^{-2}\Box_G\rho\big)\delta\rho\Big\}\\
    &\= -2\int\limits_\Mcal\diff^4x\frac{\sqrt{|G|}}{16\pi G_N}\Big\{\frac{1}{\rho^2}\big(\pa_\mu\rho\pa_\nu\rho-\sfrac12G_{\mu\nu}\pa\rho\cdot\pa\rho\big)\delta G^{\mu\nu} + \frac{2}{\rho^2m_\Kcal}(\pa\rho\cdot\pa\rho)\delta m_\Kcal\\
    &\hspace{0.45\linewidth} + \frac{2}{\rho^2}\big(\sfrac{2}{\rho}\pa\rho\cdot\pa\rho - \sfrac{2}{m_\Kcal}\pa\rho\cdot\pa m_\Kcal-\square_G\rho\big)\delta\rho\Big\}\ ,
\end{aligned}
\end{equation}
}
using \eqref{varDet} in the first and partial integration in the second line. Next, we express $S_{G_N}$ using \eqref{GNdef}:
\begin{equation}
    S_{G_N} \= 4\int\limits_\Mcal\diff^4x\frac{\sqrt{|G|}}{16\pi G_N}\(\frac{1}{\rho^2}\pa\rho\cdot\pa\rho - \frac{1}{\rho m_\Kcal}\pa\rho\cdot\pa m_\Kcal\)\ .
\end{equation}
We then compute the variation of this $S_{G_N}$ using a partial integration technique as before:
{\small
\begin{equation}\label{delSgN}
\begin{aligned}
    \delta S_{G_N} &= 4\int\limits_\Mcal\diff^4x\frac{\sqrt{|G|}}{16\pi G_N}\Big\{{-}\big(\sfrac12 G_{\mu\nu}\delta G^{\mu\nu}+G_N^{-1}\delta G_N\big)\big(\sfrac{1}{\rho^2}\pa\rho\cdot\pa\rho-\sfrac{1}{\rho m_\Kcal}\pa\rho\cdot\pa m_\Kcal\big) - \sfrac{2}{\rho^3}(\pa\rho\cdot\pa\rho)\delta\rho\\
    &\hspace{0.3\linewidth}+ \sfrac{1}{\rho^2}\pa_\mu\rho\pa_\nu\rho\delta G^{\mu\nu} + \sfrac{2}{\rho^2}\pa\rho\cdot\pa\delta\rho + \sfrac{1}{\rho m_\Kcal}\pa\rho\cdot\pa m_\Kcal\big(\sfrac{\delta\rho}{\rho}+\sfrac{\delta m_\Kcal}{m_\Kcal}\big)\\
    &\hspace{0.25\linewidth}- \sfrac{1}{\rho m_\Kcal}\pa_\mu\rho\pa_\nu m_\Kcal\delta G^{\mu\nu} - \sfrac{1}{\rho m_\Kcal}\big(\pa\delta\rho\cdot\pa m_\Kcal + \pa\rho\cdot\pa\delta m_\Kcal\big)\Big\}\\
    &= 4\int\limits_\Mcal\diff^4x\frac{\sqrt{|G|}}{16\pi G_N}\Big\{{-}\big(\sfrac12 G_{\mu\nu}\delta G^{\mu\nu}+G_N^{-1}\delta G_N\big)\big(\sfrac{1}{\rho^2}\pa\rho\cdot\pa\rho-\sfrac{1}{\rho m_\Kcal}\pa\rho\cdot\pa m_\Kcal\big) - \sfrac{2}{\rho^3}(\pa\rho\cdot\pa\rho)\delta\rho\\
    &\hspace{0.1\linewidth}+ \sfrac{1}{\rho^2}\pa_\mu\rho\pa_\nu\rho\delta G^{\mu\nu} - \sfrac{2}{\rho^2}\pa\rho\cdot\big(\sfrac{2}{m_\Kcal}\pa m_\Kcal-\sfrac{4}{\rho}\pa\rho\big)\delta\rho + \sfrac{1}{\rho m_\Kcal}\pa\rho\cdot\pa m_\Kcal\big(\sfrac{\delta\rho}{\rho}+\sfrac{\delta m_\Kcal}{m_\Kcal}\big)\\
    &\hspace{0.15\linewidth}- \sfrac{2}{\rho^2}\Box_G\rho\,\delta\rho - \sfrac{1}{\rho m_\Kcal}\pa_\mu\rho\pa_\nu m_\Kcal\delta G^{\mu\nu} + \sfrac{1}{\rho m_\Kcal}\pa m_\Kcal\cdot\big(\sfrac{1}{m_\Kcal}\pa m_\Kcal-\sfrac{3}{\rho}\pa\rho\big)\delta\rho\\
    &\hspace{0.2\linewidth}+ \sfrac{1}{\rho m_\Kcal}\pa\rho\cdot\big(\sfrac{1}{m_\Kcal}\pa m_\Kcal-\sfrac{3}{\rho}\pa\rho\big)\delta m_\Kcal + \sfrac{1}{\rho m_\Kcal}\big(\Box_G m_\Kcal\,\delta\rho+\Box_G\rho\,\delta m_\Kcal\big) \Big\}\\
    &= 4\!\!\int\limits_\Mcal\diff^4x\frac{\sqrt{|G|}}{16\pi G_N}\Big\{\frac{1}{\rho^2}\big(\pa_\mu\rho\pa_\nu\rho-\sfrac12 G_{\mu\nu}\pa\rho\cdot\pa\rho\big)\delta G^{\mu\nu} - \frac{1}{\rho m_\Kcal}\big(\pa_\mu\rho\pa_\nu m_\Kcal-\sfrac12G_{\mu\nu}\pa\rho\cdot\pa m_\Kcal\big)\delta G^{\mu\nu}\\
    &\hspace{0.1\linewidth}+ \big(\sfrac{4}{\rho^3}\pa\rho\cdot\pa\rho-\sfrac{4}{\rho^2m_\Kcal}\pa\rho\cdot\pa m_\Kcal+\sfrac{1}{\rho m_\Kcal^2}\pa m_\Kcal\cdot\pa m_\Kcal-\sfrac{2}{\rho^2}\Box_G\rho+\sfrac{1}{\rho m_\Kcal}\Box_Gm_\Kcal \big)\delta\rho\\
    &\hspace{0.3\linewidth} 
    -\big(\sfrac{1}{\rho^2m_\Kcal}\pa\rho\cdot\pa\rho-\sfrac{1}{\rho m_\Kcal}\Box_G\rho\big)\delta m_\Kcal\Big\}\ .
\end{aligned}
\end{equation}
}
In this way we obtain the variation of the $1$-loop term $\Gamma_{\ol}$ \eqref{1loopSplit} after combining the variations of $S_\Rcal$ \eqref{delSr2}, $S_T$ \eqref{delSt2}, $S_\rho$ \eqref{delSrho} and $S_{G_N}$ \eqref{delSgN} that reads
{\small
\begin{equation}\label{delGamma}
\begin{aligned}
    \delta \Gamma_{\ol} = &\int\limits_\Mcal\diff^4x\frac{\sqrt{|G|}}{16\pi G_N}\Big\{\Rcal_{\mu\nu}-\sfrac{1}{2}G_{\mu\nu}\Rcal - \sfrac12\big(T_\mu T_\nu-\sfrac12G_{\mu\nu}T\cdot T\big) - \sfrac{1}{\rho^2}\big(4\pa_\mu\rho\pa_\nu\rho - 5G_{\mu\nu}\pa\rho\cdot\pa\rho\big) \\
    &\hspace{0.1\linewidth} + \sfrac{1}{\rho m_\Kcal}\big(4\pa_\mu\rho\pa_\nu m_\Kcal - 6G_{\mu\nu}\pa\rho\cdot\pa m_\Kcal\big) - \sfrac{2}{m_\Kcal^2}\big(\pa_\mu m_\Kcal\pa_\nu m_\Kcal - G_{\mu\nu}\pa m_\Kcal\cdot\pa m_\Kcal\big) \\
    &\hspace{0.2\linewidth} -\sfrac{2}{m_\Kcal}\big(\pa_\mu\pa_\nu m_\Kcal - G_{\mu\nu} \Box_G m_\Kcal\big) + \sfrac{2}{\rho}\big(\pa_\mu\pa_\nu \rho - G_{\mu\nu} \Box_G \rho\big)\Big\} \delta G^{\mu\nu} \\
    &+\int\limits_\Mcal\diff^4x\frac{\sqrt{|G|}}{16\pi G_N}\Big\{{-2}\Rcal - T\cdot T + \sfrac{8}{\rho^2}\pa\rho\cdot\pa\rho - \sfrac{8}{\rho m_\Kcal}\pa\rho\cdot\pa m_\Kcal + \sfrac{4}{m_\Kcal^2}\pa m_\Kcal\cdot\pa m_\Kcal\\
    &\hspace{0.3\linewidth}- \sfrac{4}{\rho}\Box_G\rho+\sfrac{4}{m_\Kcal}\Box_G m_\Kcal \Big\}\rho^{-1}\delta\rho \\
    &+\int\limits_\Mcal\diff^4x\frac{\sqrt{|G|}}{16\pi G_N}\Big(2\Rcal + T\cdot T - \sfrac{8}{\rho^2}\pa\rho\cdot\pa\rho+\sfrac{4}{\rho}\Box_G\rho \Big)m_\Kcal^{-1}\delta m_\Kcal \\
    &- \int\limits_\Mcal\diff^4x\frac{\rho^2}{16\pi G_N}\varepsilon^{\nu\sigma\mu\kappa}\Big\{T_\kappa T_{\nu\mu}^{\ \ \adot} + m_\Kcal^{-2}E^\adot_{\mu}\pa_\nu(m_\Kcal^2T_\kappa)\Big\}\delta E_{\adot\sigma}\ .
\end{aligned}
\end{equation}
}

\section{Expressing $\delta G^{\mu\nu}$ and $\delta\sigma$ in terms of $\delta E_{\adot\mu}$}
\label{VarTransf}
\noindent
First of all, we can relate the determinants of effective metric $G_{\mu\nu}$ and the frame $E_{\adot\mu}$ using the relations \eqref{auxMetric} and \eqref{confMetric},
\begin{equation}
    \sqrt{|G_{\mu\nu}|} \= \rho^4|E_{\adot\mu}|\ .
\end{equation}
This allows to express the dilaton $\rho$ in terms of the frame using \eqref{rhoDef} as
\begin{equation}
    \rho^2 \= \frac{\rho_{_\Mcal}}{|E_{\adot\mu}|}\ ,
\end{equation}
whose variation can then be obtained by noting that $\rho_{_\Mcal}$ is rigid:
\begin{equation}
\begin{aligned}
    \delta\sigma\ :=\ \rho^{-1}\delta\rho &\= \frac{1}{2\rho^2}\delta(\rho^2) \= \frac{1}{2\rho^2}\delta\(\frac{\rho_{_\Mcal}}{|E_{\adot\mu}|}\)\\
    &\= -\frac{1}{2\rho^2}\frac{\rho_{_\Mcal}}{|E_{\adot\mu}|^2}\delta|E_{\adot\mu}|\\
    &\= -\frac{1}{2\rho^2}\frac{\rho_{_\Mcal}}{|E_{\adot\mu}|^2}|E_{\adot\mu}|E^{\adot\sigma}\delta E_{\adot\sigma}
    \= -\frac{1}{2}E^{\adot\sigma}\delta E_{\adot\sigma}\ .
\end{aligned}
\end{equation}
With this result in hand, we can compute the variation of the inverse metric $G^{\mu\nu}$ \eqref{confMetric} while also using \eqref{auxMetric} and symmetry of indices as follows:
\begin{equation}
\begin{aligned}
    \delta G^{\mu\nu} &\= \delta\(\rho^{-2}\gamma^{\mu\nu}\) \= \delta\(\rho^{-2}E^\mu_\adot E^{\adot\nu}\)\\
    &\= -\frac{2}{\rho^3}\gamma^{\mu\nu}\delta\rho + \frac{2}{\rho^2}E^\mu_\adot\delta E^{\adot\nu} \\
    &\= -\frac{2}{\rho^2}\gamma^{\mu\nu}\(-\frac{1}{2}E^{\adot\sigma}\delta E_{\adot\sigma}\) + \frac{2}{\rho^2}E^\mu_\adot\delta E^{\adot\nu}\\
    &\= G^{\mu\nu}E^{\adot\sigma}\delta E_{\adot\sigma} + 2\rho^{-2}E^\mu_\adot\delta E^{\adot\nu}\ . 
\end{aligned}
\end{equation}

\section{Divergence constraint}
\label{sec:div-einstein}
\noindent
Taking the divergence $\nabla^\mu$ of 
\eqref{Einstein-e-mod4} and using the conservation of the Einstein tensor and of the energy-momentum tensor of matter, we obtain
\begin{equation}
\begin{aligned}
    0 &\= 32\pi \nabla^\mu{G_N} \Big(C_{\mu\lambda} + 
    E^\adot_\lambda\partial_\mu C_\adot
    -\frac 12 G_{\mu\lambda}(C + \rho^{-2} E^{\adot\mu}\partial_\mu C_\adot)\Big)
    + 2\Box_G\sigma\pa_\lambda\sigma 
    - 2\pa_\lambda(\pa\sigma\cdot\pa\sigma) 
\end{aligned}
\end{equation}
dropping the vacuum energy contributions for simplicity, which is justified at late times.
The axionic vector field $T_\mu$ is dropped for simplicity. This constraint can  be used to fix the 
remaining ambiguity in 
$C_{\adot\nu}\to C_{\adot\nu} + \pa_\nu C_\adot$, which determines precisely the 4 degrees of freedom 
in $C_\adot$. In that sense, the divergence constraint of the Einstein equations allows us to uniquely determine the anharmonicity tensor $C_{\mu\nu}$ which arises from the Yang-Mills background via \eqref{Camu-def-BG}. Therefore our modified Einstein equations uniquely determine the geometry of space-time for given matter, dilaton, and axionic field $T_\mu$.

\section{Scalar components}
\noindent
There are two equations for the scalar components, one from 
\eqref{Einstein-e-mod2-trace}:
{\small 
\begin{equation}
\begin{aligned}
    \Rcal
    &\= {8\pi G_N}\bigg[-  T^{(m)}
    + 4\rho^{-4}\big(2\rho^2F_{\Kcal}^2m_\Kcal^4
    -C_1m_\Kcal^4+\frac{C_2}{R^4}+3\frac{C_3}{R^8m_\Kcal^4}\big) 
    + 4 C\bigg] 
    + 10\pa\sigma\cdot\pa\sigma 
    - 6 \Box_G\sigma  
\end{aligned}
\end{equation}
}
and one  from the eom for $\lambda$ \eqref{eomMass}
\begin{equation}
    \Rcal + \frac{1}{2}T\cdot T - 2\pa\sigma\cdot\pa\sigma + 2\Box_G\sigma 
    - 32\pi G_N \rho^{-2}F_\Kcal^2 m_\Kcal^4 + 32\pi G_N \rho^{-4}
    \big( C_1 m_\Kcal^4 
    - C_3 \frac{1}{R^8 m_\Kcal^4}\big)
    \ = 0\ .
\end{equation}
We can eliminate the Ricci scalar from these equations to obtain the following relation among various non-geometric quantities
\begin{equation}
    \frac{1}{2}T\cdot T + 8\pa\sigma\cdot\pa\sigma - 4\Box_G\sigma 
    - {8\pi G_N}T^{(m)}
    + {32\pi G_N\rho^{-4}}\(\rho^2F_{\Kcal}^2m_\Kcal^4
    +\frac{C_2}{R^4}+2\frac{C_3}{R^8m_\Kcal^4} 
    + 4 C\) \= 0\ .
\end{equation}
The significance of this equation is not clear.

\end{document}